# Storing a Trie with Compact and Predictable Space

DONG Yuxuan <https://www.dyx.name>

This paper proposed a storing approach for trie structures, called *coordinate hash trie*. The basic idea is using a global hash table with a special hash function to store all edges of a trie. For a trie with $n$ nodes and an alphabet with size $m$, the execution time of finding, inserting and deleting a child node, is $O(1)$ for the average case, $O(m)$ for the worst case. The space used by this approach is $O(n)$, unrelated to $m$. The constant of space consumption is predictable, with no need for reallocation or resizing. In addition, this approach is very easy to implement.

**Table of Contents**

---

## Introduction

The problem of searching, inserting, and deleting a string in a string set, arises frequently in programming. Trie [4] is a widely used data structure for this problem.

A trie is a tree for storing a set of strings. Each edge of a trie is labeled with a symbol in the alphabet. Figure 1 shows a trie of strings {`he`, `she`, `his`, `hers`}, which were inserted in order. Node `0` is the root node. A double circle node denotes a *terminal* node where a string terminates.

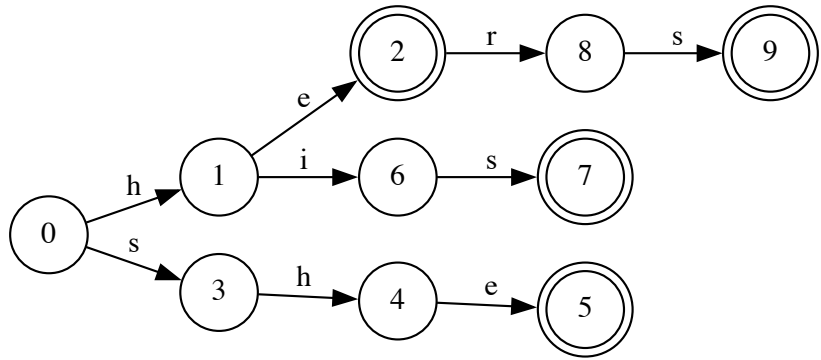

Figure1. A trie of {he, she, his, hers}

A trie has three basic operations.

- Node walking: Given a node $x$ and symbol $y$, find the child node $z$, which is reached by the symbol $y$;

- Node insertion: Given a node $x$ and symbol $y$, create a new child node $z$ of $x$ and connect $x$ and $z$ with the symbol $y$;

- Node deletion: Given a node $x$ and symbol $y$, delete the child leaf node $z$ of $x$ which is connected with the symbol $y$.

Suppose there are $n$ nodes in a trie, and the size of the alphabet is $m$. If the execution time of basic operations is $f(n,m)$, then the execution time of searching, inserting, and deleting a string with length $l$ would all be $O(lf(n,m))$.

The most straightforward and time-efficient implementation of trie is using a direct-mapped array for each node to store its children. This approach is called the *direct-mapped trie* in this paper. A direct-mapped trie can be represented by a two-dimension array $A[n][m]$. If there is an edge from $i$ to $k$, labeled with the symbol $j$, we will have $A[i][j] = k$. If there is no such edge, $A[i][j]$ is set to 0. This works because there is no edge pointing to the root node.

The direct-mapped approach has $f(n,m) = O(1)$, thus it's very efficient. The matrix $A$, however, takes $O(nm)$ space. This makes the direct-mapped trie unpractical for large alphabets or devices with a restricted primary memory. The matrix $A$ is usually very sparse. For example, in figure 1 with the alphabet {a,b,c,...,z}, there are 26 columns in each row in $A$, but a row contains at most two valid elements. The sparse nature of $A$ gives us opportunities for compression.

Several efforts were made to compress a trie. However, some of these approaches significantly increase the execution time of one or more basic operations. Others make the actual space consumption hard to estimate.

This paper proposed an approach for storing a trie. The execution time of basic operations is $O(1)$ for the average case, $O(m)$ for the worst case. The space consumption is $O(n)$. Because our approach requires no resizing or reallocation, the actual space consumption is predictable. Thus it's practical for large alphabets and devices with a restricted primary memory.

## Our Approach

For a trie with $n$ nodes and the alphabet with size $m$. We use an integer $x \in [0, n)$ to denote a node, and an integer $y \in [0, m)$ to denote a symbol in the alphabet.

Our approach, called *coordinate hash trie*, uses a global hash table, called the *edge table*, to store all edges in the trie. The edge table represents a key-value dictionary. The edge from node $x$ to node $z$, labeled with $y$, is represented by a dictionary item $(x, y) \to z$ in the edge table. $(x, y)$ is called the *edge key*. $z$ is called the *edge value*.

The hash function used in our approach is

$$h(x, y) = (xm + y) \bmod H,$$

where $H$ is the number of slots in the edge table.

The trie contains $n - 1$ edges. Thus we could take $H = (n - 1)/\alpha$, where $\alpha$ is a positive real number constant, known as the load factor of the hash table.

Basic operations of trie thus become search, insertion, and deletion operations of hash table.

At the first glance, accessing an item in the edge table may search the whole table for the worst case. However, we will prove that, with the above hash function, the execution time could be bound to $O(m)$ for the worst case.

## Complexity Analysis

The space consumption of a coordinate hash tire is clearly $O(n)$. The execution time of basic operations of a coordinate hash trie depends on how the hash table is implemented. We assume an implementation meets the following conditions.

**Assumption 1** The average execution time of searching, inserting, and deleting an item in the hash table is $O(1)$;

**Assumption 2** The worst execution time of searching, inserting, and deleting an item keyed with $k$ in the hash table, is at most proportional to the number of keys, which have the same hash value as $k$ has.

Most implementations of hash table meet or approximately meet assumption 1 and 2.

**Theorem 1** The average execution time of node walking, node insertion, and node deletion of a coordinate hash trie is $O(1)$.

**Proof** Straightforward from assumption 1.

□

To give the worst execution time, we need to figure out, for a given edge key $(x, y)$, how many edge keys have the same hash value as $(x, y)$ has.

We give a definition for convenience first.

**Definition 1** The *GCD coordinate* of an edge key $(x, y)$ is a tuple $(x', y')$ which meets:

$$\begin{cases} x' = \lfloor (xm + y)/gcd(H, m) \rfloor, \\ y' = (xm + y) \bmod gcd(H, m), \end{cases}$$

where we use $gcd(a, b)$ to denote the greatest common divider of non-negative integers $a$ and $b$.

Edge keys and their GCD coordinates are one-to-one mapped. In addition, for each edge key $(x, y)$ and its GCD coordinate $(x', y')$, we have

$$\begin{cases} xm + y = x' gcd(H, m) + y', \\ 0 \leq x' < nm/gcd(H, m), \\ 0 \leq y' < gcd(H, m). \end{cases}$$

Using the concept of GCD coordinates, we could give the condition that two edge keys have the same hash value.

**Lemma 1** Given two edge keys $(x_1, y_1)$ and $(x_2, y_2)$, $h(x_1, y_2) = h(x_2, y_2)$ if and only if

$$\begin{cases} x'_1 \equiv x'_2 \pmod{H/gcd(H, m)}, \\ y'_1 = y'_2, \end{cases}$$

where $(x'_i, y'_i)$ is the GCD coordinate of $(x_i, y_i)$, for $i = 1, 2$.

**Proof** Take $G = gcd(H, m)$.

$$\begin{aligned}
& h(x_1, y_1) = h(x_2, y_2) \\
\Longrightarrow \quad & x_1 m + y_1 \equiv x_2 m + y_2 \pmod{H} \\
\Longrightarrow \quad & x'_1 G + y'_1 \equiv x'_2 G + y'_2 \pmod{H} \\
\Longrightarrow \quad & (x'_1 - x'_2) G + (y'_1 - y'_2) \equiv 0 \pmod{H} \\
\Longrightarrow \quad & (x'_1 - x'_2) G + (y'_1 - y'_2) \text{ is a multiple of } H \\
\Longrightarrow \quad & (x'_1 - x'_2) G + (y'_1 - y'_2) \text{ is a multiple of } G \\
\Longrightarrow \quad & y'_1 - y'_2 \text{ is a multiple of } G.
\end{aligned}$$

Because $|y'_1 - y'_2| < G$, we have $y'_1 - y'_2 = 0$. Thus we have $y'_1 = y'_2$.

Then we get

$$\begin{aligned}
& x'_1 G \equiv x'_2 G \pmod{H} \\
\Longrightarrow \quad & x'_1 \equiv x'_2 \pmod{H/G}.
\end{aligned}$$

Another direction of the proof is similar.

□

**Theorem 2** The worst execution time of node walking, node insertion, and node deletion of a coordinate hash trie is at most proportional to $\lceil \alpha m \rceil$.

**Proof** For a given edge key $(x_0, y_0)$ with the GCD coordinate $(x'_0, y'_0)$, we denote the number of edge keys which has the same hash value with $(x_0, y_0)$ to be $t$.

According to assumption 2, the worst execution time of node walking, node insertion, and node deletion in the coordinate hash trie is at most proportional to $t$.

According to lemma 1, $t$ is equal to the number of $x'$ which meet

$$x' \equiv x'_0 \pmod{\frac{H}{\gcd(H, m)}},$$

where $x'$ is the x-component of the GCD coordinate of an edge key.

Because $0 \leq x' < nm/\gcd(H, m)$, there are at most $nm/\gcd(H, m)$ possible values of $x'$. Thus we have

$$t \leq \left\lceil \frac{nm/\gcd(H, m)}{H/\gcd(H, m)} \right\rceil = \lceil \alpha m \rceil.$$

□

# Related Work

A general approach for compressing a sparse matrix, called *row displacement*, was proposed by [1], [6], and analyzed by [5].

An approach based on row displacement, known as *double-array trie*, was proposed by [2]. Double-array trie keeps $O(1)$ worst execution time of node walking and deletion. However, the tight bound of the space used by a double-array trie is hard to estimate by parameters $n$ and $m$. In addition, the worst execution time of node insertion is significantly increased. Under the assumptions that a double array trie uses $O(n + cm)$ space, where $c$ is a constant, the worst execution time of node insertion is $O(nm + cm^2)$.

Another approach is using $n$ binary search trees, each for a node, to store the children of the node. This approach reduces the space of a trie to $O(n)$. However, the execution time of basic operations is increased to $O(\log m)$.

We could also use $n$ hash tables, each for a node, to store the children of the node. This approach reduces the space of a trie to $O(n)$. This approach also gives $O(1)$ execution time of the basic operations for the average case, and $O(m)$ for the worst case. However, this approach installs $n$ hash tables with different sizes. Each hash table requires an initial capacity. If the initial capacities are too large, there will be space waste. If the capacities are insufficient, resizing and reallocation must be made. This makes the actual space used by a trie hard to estimate by parameters $n$ and $m$. In addition, the resizing and reallocation can significantly affect the execution time if they occur frequently.

## Remarks

A C implementation of the coordinate hash trie is provided online: <https://github.com/dongyx/chtrie>.

Our approach can be generalized to store any sparse matrix. An approach for sparse matrix storage using a similar idea was proposed by [3], but it didn't give a theoretical analysis.